\documentclass[12 pt]{article}
\usepackage[utf8x]{inputenc}
\usepackage{amsmath}
\usepackage{amsfonts}
\usepackage{graphicx}
\usepackage{empheq}
\usepackage[sort&compress,numbers]{natbib}
\usepackage{doi}
\hoffset= - 0.9in         
\newcommand*\widefbox[1]{\fbox{\hspace{2em}#1\hspace{2em}}}
\newcommand{\p}{\partial}
\newcommand{\f}{\frac}
\newcommand{\s}{\sqrt}
\newcommand{\wh}{\widehat}

\newcommand{\e}{\epsilon}
\newcommand{\be}{\beta}
\newcommand{\al}{\alpha}

\newcommand{\cf}{\mathcal{F}}

\newcommand{\co}{\mathcal{O}}
\newcommand{\cm}{\mathcal{M}}

\newcommand{\nn}{\nonumber}

\usepackage{hyperref}
\usepackage{tikz}
\usepackage{epigraph}
\usepackage{float}

\title{Series expansion of fusion kernel} 
 
\begin{document}
 \title{ On fusion kernel in Liouville theory}
 \author{{ N.Nemkov}\thanks{{\small
{\it Institute for Theoretical and Experimental Physics (ITEP), Moscow, Russia} and {\it Moscow Institute of Physics and Technology (MIPT), Dolgoprudny, Russia}; nnemkov@gmail.com} }
\date{ }}
\date{\today}
\maketitle
\vspace{-5.0cm}

\begin{center}
 \hfill ITEP/TH-21/14\\
\end{center}

\vspace{3.5cm}
\begin{abstract}
We study fusion kernel for non-degenerate conformal blocks in Liouville theory as a solution to the difference equations originating from the pentagon identity. We suggest an approach to these equations based on 'non-perturbative' series expansion which allows to calculate the fusion kernel iteratively. We also find exact solutions for the cases when the central charge is $c=1+6(b-b^{-1})^2$ and  $b~\in \mathbb{N}$. For   $c = 1$ our result reproduces the formula, obtained earlier from
analytical continuation via Painlev\'e equation. However, in our case it appears in a significantly simplified form.

\end{abstract}
\newpage
 \tableofcontents
\newpage
\section{Introduction}
Conformal blocks are the main special functions of conformal theories \cite{BPZ}. CFTs arise in numerous areas of the modern theoretical physics and have for a long time been under indiverted attention. Renewed interest have been caused by the celebrated AGT correspondence \cite{AGT} relating 2d CFTs to higher-dimensional SUSY gauge theories. From this perspective conformal blocks of 2d CFTs become partition functions of gauge theories.

Many important properties of conformal blocks are discovered, but still there is a long way to go. Present paper addresses one particular aspect of their behaviour. On the space of conformal blocks naturally acts the fusion algebra \cite{MS}. On the gauge theory side fusion transformations correspond to duality transformations. In certain cases called degenerate the space of conformal blocks is of finite dimension and constituents of the fusion algebra are just finite matrices. These matrices are known explicitly or can be straightforwardly calculated, at least in principle. However, in a generic situation the space of conformal blocks is infinite-dimensional and extensions of these finite-dimensional matrices are not fully understood.

The problem of finding the fusion kernel in the Liouville theory for generic values of parameters was initiated and solved in \cite{PT1,PT2}. There remain doubts whether the results obtained in \cite{PT1,PT2} are applicable to the special case of the unit central charge.  This special case was handled in \cite{Iorgov}. Nominally, the results of \cite{PT1,PT2,Iorgov}  provide the answer to the question. However, manifest formulas obtained there are very sophisticated and may cover some important properties and/or admit remarkable simplification. In this paper we take yet another approach to the problem which hopefully helps to clear up a matter.

We derive certain difference equations on the fusion kernel. With the help of some ansatz, these equations can be solved iteratively giving explicit expansion of the generic fusion kernel. 
Expansion is made in terms of 'non-perturbative' parameters controlling large intermediate dimension limit. In the case of the unit central charge (and also for $c=1+6(b-b^{-1})^2, b\in \mathbb{N}$) we find the exact solution, which appears to be a rather simplified version of the expression for the $c=1$  fusion kernel suggested in \cite{Iorgov}.  
\section{Notation}

 There is some diversity of notation in the literature and we start by declaring ours.
 
 Central charge of the conformal theory is denoted by $c$ and parametrized as
\begin{eqnarray}
c=1-6Q^2,\quad Q=b-b^{-1}\label{central charge}
\end{eqnarray}
Our most decorated notation for conformal block is 

\begin{eqnarray}
B_a\begin{bmatrix}
a_2&a_3\\a_1 &a_4
\end{bmatrix}(x)
\end{eqnarray}
Or, graphically
 
\begin{figure}[H]
\centering
\begin{tikzpicture}

 \node at (-1,1) {$B_a\begin{bmatrix}
 a_2&a_3\\a_1 &a_4
 \end{bmatrix}(x)\quad=$};
 \draw[line width=2pt] (1,0.3) node [below] {$a_1,0$} -- ++(0.5,0)  -- ++(0.5,0) -- ++(0,1.5) node[left] {$a_2,x$} -- ++(0,-1.5) -- ++(0.5,0) node [below] {$a$} -- ++(0.5,0) -- ++(0,1.5) node[right]{$a_3,1$} -- ++(0,-1.5) -- ++ (1,0) node [below] {$a_4,\infty$};
\draw +(6,0.75); 
\end{tikzpicture}
\end{figure}\label{CBs} 
 \noindent Here $a$ and $a_i=\{a_1,a_2,a_3,a_4\}$ are internal and external Liouville momenta respectively. They parametrize conformal dimensions $\Delta,\Delta_i$ as
\begin{eqnarray}
\Delta=a(Q-a),\quad \Delta_i=a_i(Q-a_i) \label{dimension par}
\end{eqnarray}
Parameter $x$ has an interpretation of the position of the field with momentum $a_2$ if the other fields are placed at points $0,1,\infty$. Fusion kernel is denoted by
\begin{eqnarray}
\cf_{aa'}\begin{bmatrix}
a_2&a_3\\a_1&a_4
\end{bmatrix}
\end{eqnarray}
Often we do not manifestly depict dependence on external momenta and write simply
\begin{eqnarray}
B_a(x),\qquad\cf_{aa'}
\end{eqnarray}
for conformal block and fusion kernel respectively.

\section{Fusion algebra}
Conformal blocks naturally appear in the decomposion of correlation functions in CFTs \cite{BPZ}. This decomposition can be performed in different ways giving rise to different conformal blocks. Requirement of the decompositions equivalence imposes relations between different types of conformal blocks \cite{MS}. 

Along with conformal block \ref{CBs} which is often called $s-$channel we consider $t-$channel conformal block defined pictorially by
\begin{figure}[H]
\centering
\begin{tikzpicture}
 \node at (-1,0.7) {$B^{t}_a\begin{bmatrix}
 a_2&a_3\\a_1&a_4
 \end{bmatrix}(x)\quad=$};
\draw[line width=2pt] (1,0)node [below] {$a_1,0$} -- ++(1.5,0)--++(1.5,0) node[below]{$a_4,\infty$}--++(-1.5,0)-- ++(0,0.75) node[below right]{$a$} -- ++(-0.75,0.75)node[left]{$a_2,x$} -- ++(0.75,-0.75)--++(0.75,0.75) node[right]{$a_3,1$};
\end{tikzpicture}
\end{figure}

$s-$ and $t-$channel conformal blocks are believed to be related by a linear integral transformation
\begin{eqnarray}
 B^{s}_a\begin{bmatrix}
  a_2&a_3\\a_1&a_4
  \end{bmatrix}(x)=\int da'\cf_{aa'}\begin{bmatrix}
  a_2&a_3\\a_1&a_4
  \end{bmatrix}\,B^{t}_{a'}\begin{bmatrix}
   a_2&a_1\\a_3&a_4
   \end{bmatrix}(x)
\end{eqnarray}
\noindent or in terms of graphs (here and further we suppress coordinate labels)

\begin{figure}[H]
\centering
\begin{tikzpicture}
\draw[line width=2pt] (0,0) node [below] {$a_1$} -- ++(0.5,0)  -- ++(0.5,0) -- ++(0,1.5) node[left] {$a_2$} -- ++(0,-1.5) -- ++(0.5,0) node [below] {$a$} -- ++(0.5,0) -- ++(0,1.5) node[right]{$a_3$} -- ++(0,-1.5) -- ++ (1,0) node [below] {$a_4$};
\draw +(6,0.75) node{$=\int  da'\,\cf_{aa'}\begin{bmatrix} a_2 & a_3\\a_1 & a_4\end{bmatrix}$};
\draw[line width=2pt] ++(9,0)node [below] {$a_1$} -- ++(1.5,0)--++(1.5,0) node[below]{$a_4$}--++(-1.5,0)-- ++(0,0.75) node[below right]{$a'$} -- ++(-0.75,0.75)node[left]{$a_2$} -- ++(0.75,-0.75)--++(0.75,0.75) node[right]{$a_3$};
\end{tikzpicture} 
\end{figure}

 \noindent This transformation is called the fusion transformation and $\cf_{aa'}$ is called the fusion kernel. 
 
 There is a special case when one of the fields in a conformal block is degenerate. Degeneration of a field corresponds to a special value of its dimension/momentum. In this paper we use only degenerate fields with momentum 
 \begin{eqnarray}
 a_\text{deg}=-b/2
 \end{eqnarray}

 Consider conformal block with $a_2=-b/2$. It an be shown that in this case internal momenta $a,a'$ can no longer be arbitrary but only take two possible values 
 \begin{eqnarray}
 a=a_1\pm b/2,\quad a'=a_3\pm b/2
 \end{eqnarray}
 
 So the space of conformal blocks with fixed external momenta is two-dimensional while the corresponding fusion kernel is just $2\times2$ matrix. In a suitable normalization of conformal blocks this fusion matrix is \cite{PT1}
 \begin{eqnarray}
\boxed{\cf_{a_1+s_1b/2,a_3+s_2b/2}\begin{bmatrix}-b/2 & a_3\\ a_1 & a_4\end{bmatrix}\equiv \cf_{s_1,s_2}\begin{bmatrix}-b/2 & a_3\\ a_1 & a_4\end{bmatrix} =s_1\f{\sin{\pi b(a_4+s_1 a_3-s_2 a_1 -(1+s_1-s_2)b/2))}}{\sin{\pi b(2a_3-b)}}}\label{Fa2}
 \end{eqnarray} 
 where $s_1,s_2=\pm$.
 
 These matrices are related to the standard ones $F_{aa'}$ by 
 \begin{eqnarray}
 \cf_{aa'}\begin{bmatrix}
 a_2&a_3\\a_1&a_4
 \end{bmatrix}=\frac{N_a\begin{bmatrix}
  a_2&a_3\\a_1&a_4
  \end{bmatrix}}{N_{a'}\begin{bmatrix}
   a_2&a_1\\a_3&a_4
   \end{bmatrix}} F_{aa'}\begin{bmatrix}
  a_2&a_3\\a_1&a_4
  \end{bmatrix}\label{fmren}
 \end{eqnarray}
with
\begin{align}
&N_a\begin{bmatrix}
  a_2&a_3\\a_1&a_4
  \end{bmatrix}=V(a_4;a_3,a)V(a;a_2,a_1)\nn\\
  &V(a_1;a_2,a_3)=\nn\\&\frac{\Gamma_b(a_1+a_2+a_3+b-2Q)\Gamma_b(a_1-a_2-a_3+b)\Gamma_b(-a_1+a_2-a_3+b)\Gamma_b(-a_1-a_2+a_3+b)}{\Gamma_b(2a_1+b-2Q)\Gamma_b(-2a_2+b)\Gamma_b(-2a_3+b)}\label{norm}
\end{align}
 Here $\Gamma_b(x)$ is related to the double gamma function
 \begin{eqnarray}
 \Gamma_b(x)=\Gamma_2(x|b,b^{-1})
 \end{eqnarray}  
 and satisfies
 \begin{eqnarray}
 \frac{\Gamma_b(x+b)}{\Gamma_b(x)}= \frac{\s{2\pi}b^{bx-1/2}}{\Gamma(bx)}
 \end{eqnarray}

  Fusion kernel $\cf_{aa'}$ must obey certain consistency requirements. In the next section we turn these requirements, together with the known form of the fusion matrix for the \textit{degenerate} case \eqref{Fa2}, into a set of difference equations on the fusion kernel with \emph{arbitrary} values of momenta.
  \section{Difference equations on fusion kernel}
  Consider two paths leading from the upper left state ($s$-channel conformal block with a degenerate insertion at the $a_1$ leg) to the lower right state ($t$-channel conformal block with a degenerate insertion at the $a_2$ leg)
  \begin{figure}[H]
  \centering
  \begin{tikzpicture}[scale=1]
  
  \draw[line width=2pt] (0,0) node [below] {$a_{s_1}$} -- ++(0.5,0)  -- ++(0.5,0) -- ++(0,1.5) node[left] {$a_2$} -- ++(0,-1.5) -- ++(0.5,0) node [below] {$a$} -- ++(0.5,0) -- ++(0,1.5) node[right]{$a_3$} -- ++(0,-1.5) -- ++ (1,0) node [below] {$a_4$};
  \draw[line width=2pt,dashed, red] (.5,0) -- (.5,.75);
  \node[below] at (0.75,-0.07){$a_1$};
  
  \draw +(6,0.75) node{$=\int da'\,\cf_{aa'}\begin{bmatrix} a_2 & a_3\\a_1 & a_4\end{bmatrix} $};
  
  \draw[line width=2pt] ++(9,0)node [below] {$a_{s_1}$} -- ++(1.5,0)--++(1.5,0) node[below]{$a_4$}--++(-1.5,0)-- ++(0,0.75) node[below right]{$a'$} -- ++(-0.75,0.75)node[left]{$a_2$} -- ++(0.75,-0.75)--++(0.75,0.75) node[right]{$a_3$};
  
  \draw[line width=2pt,dashed, red] (9.5,0) -- (9.5,.75);
  
  \node[below] at (10,-0.07){$a_1$};
  
  \draw (1.45,-1.6)--(1.45,-1.9);
  \draw (1.55,-1.6)--(1.55,-1.9) node[below ]{$\sum\limits_{s_2}\co_A^{s_1,s_2}(a_i,a)$};
  
  \draw[line width=2pt] (0,-5) node [below] {$a_{s_1}$} -- ++(0.5,0)  -- ++(0.5,0) -- ++(0,1.5) node[left] {$a_2$} -- ++(0,-1.5) -- ++(0.5,0) node [below] {$a$} -- ++(0.5,0) -- ++(0,1.5) node[right]{$a_3$} -- ++(0,-1.5) -- ++ (1,0) node [below] {$a_4$};
  \draw[line width=2pt,dashed,red] (1,-4)--+(-0.75,0);
  \node[left] at (1,-4.5){$a_{s_2}$}; 
  \node at (6,-4.25){$=\int da'\,\cf_{aa'} \begin{bmatrix}  a_{s_2} & a_3\\a_{s_1} & a_4\end{bmatrix}$};
  
  \draw[line width=2pt] ++(9,-5)node [below] {$a_{s_1}$} -- ++(1.5,0)--++(1.5,0) node[below]{$a_4$}--++(-1.5,0)-- ++(0,0.75) node[below right]{$a'$} -- ++(-0.75,0.75)node[left]{$a_2$} -- ++(0.75,-0.75)--++(0.75,0.75) node[right]{$a_3$};

  \draw[line width=2pt,dashed, red] (10,-5+1.25) -- +(-0.5,-.5);
  \node at (10.1,-5+.8) {$a_{s_2}$};
  
  \draw (9+1.45,-1.6)--(9+1.45,-1.9);
  \draw (9+1.55,-1.6)--(9+1.55,-1.9) node[below]{$\sum\limits_{s,s_2}\co_B^{s;s_1,s_2}(a_i,a')e^{s\f{b}2\p_{a'}}$};
  \end{tikzpicture}
  \label{two paths}
  \end{figure}
  
  Let us explain the picture. Red discrete lines depict  insertions of a degenerate fields with momentum $-b/2$. Momentum of the leg to which degenerate field is attached gets shifted by $\pm b/2$ and labels $a_{s_1}\equiv a_1+  s_1 b/2, a_{s_2}\equiv a_2+ s_2 b/2$ with $s_1,s_2=\pm$ reflect precisely that.
  
   Horizontal equalities in the upper and in the lower row are simply definitions of the fusion transformation. Vertical equality in the first column is also a manifest fusion transformation applied to the subgraph with external legs of dimensions $\{a_{s_1},-b/2,a_2,a\}$. Respectively, coefficients $ \co_A^{s_1,s_2}(a_i,a)$ are
   \begin{eqnarray}
\boxed{    \co_A^{s_1,s_2}(a_i,a)=\cf_{-s_1,s_2}\begin{bmatrix}
   -b/2&a_2\\a_{s_1}&a
   \end{bmatrix}}
   \end{eqnarray}

   Finally, vertical equality in the second column is obtained by applying fusion transformation twice. First move transports degenerate insertion to the intermediate channel and is performed by matrix 
   \begin{eqnarray}
   \cf_{-s_1,s}\begin{bmatrix}
   -b/2&a'\\a_{s_1}&a_4
   \end{bmatrix}
   \end{eqnarray}

   Second move transports degenerate insertion further to 	$a_2$ leg and introduces matrix
   \begin{eqnarray}
   \cf_{-s,s_2}^{-1}\begin{bmatrix}
   a_2&a_3\\-b/2&a'_s\end{bmatrix}=\cf_{-s,s_2} \begin{bmatrix}
      -b/2&a_2\\a'_s&a_3\end{bmatrix}
   \end{eqnarray}
   Thus, coefficients $\co_B^{s;s_1,s_2}$ are
   \begin{eqnarray}
\boxed{   \co_B^{s;s_1,s_2}=   \cf_{-s_1,s}\begin{bmatrix}
      -b/2&a'\\a_{s_1}&a_4\end{bmatrix}\cf_{-s,s_2} \begin{bmatrix}
            -b/2&a_2\\a'_s&a_3\end{bmatrix}}
   \end{eqnarray}
      In the process of transporting degenerate field along internal leg internal momentum $a'$ gets shifted by $\pm b/2, a'_s=a'\pm sb/2, s=\pm$, which explains appearance of shift operators $e^{s\frac{b}{2}\p_{a'}}$.  
      
      Thereby, equivalence of the two paths produces equation 
      \begin{multline}
      \sum_{s_2}\co_A^{s_1,s_2}(a_i,a)\int da' \cf_{aa'}\begin{bmatrix}
      a_{s_2}&a_3\\a_{s_1}&a_4
      \end{bmatrix} B^{5-\text{point}}_{a'}(x)=\\       \int da' \cf_{aa'}\begin{bmatrix}
      a_2&a_3\\a_1&a_4
      \end{bmatrix} \sum_{s,s_2}\co_B^{s;s_1,s_2}(a_i,a')e^{s\f{b}2\p_{a'}}  B^{5-\text{point}}_{a'}(x)\label{diff int}
      \end{multline}
      Here $B^{5-\text{point}}_{a'}(x)$ stands for the four point conformal block with the degenerate insertion at the $a_2$ leg, the destination point at the picture.
      
      One can make a shift of variables in the integral on the second line of \eqref{diff int} to remove the shift operators from the conformal block
       \begin{multline}
             \sum_{s_2}\co_A^{s_1,s_2}(a_i,a)\int da' \cf_{aa'}\begin{bmatrix}
             a_{s_2}&a_3\\a_{s_1}&a_4
             \end{bmatrix} B^{5-\text{point}}_{a'}(x)=\\       \int da' \left(e^{-s\f{b}2\p_{a'}}\left(\cf_{aa'}\begin{bmatrix}
             a_2&a_3\\a_1&a_4
             \end{bmatrix} \sum_{s,s_2}\co_B^{s;s_1,s_2}(a_i,a')\right)\right)  B^{5-\text{point}}_{a'}(x)\label{diff int 2}
             \end{multline}
             
             Let us introduce operators
              
 \begin{empheq}[box=\widefbox]{align}
  \hat{\co}_A^{s_1,s_2}(a_i,a)&=\co_{A}^{s_1,s_2}(a_i,a)e^{\f{b}2(s_1\p_{a_1}+s_2\p_{a_2})} \nn\\ \hat{\co}_B^{s_1,s_2}(a_i,a')&=\sum_{s}e^{-s\f{b}2\p_{a'}}\co_B^{s;s_1,s_2}(a_i,a') 
 \end{empheq} 
              
             Then, \eqref{diff int 2} is rewritten as
             \begin{multline}
             \int da'\, \sum_{s_2}\left(\hat{\co}_A^{s_1,s_2}(a_i,a)\cf_{aa'}\begin{bmatrix}
             a_2&a_3\\a_1&a_4\end{bmatrix} \right)
             B^{5-\text{point}}_{a'}(x)=\\\int da'\, \sum_{s_2}\left(\hat{\co}_B^{s_1,s_2}(a_i,a)\cf_{aa'}\begin{bmatrix}
                          a_2&a_3\\a_1&a_4 
                          \end{bmatrix}\right)B^{5-\text{point}}_{a'}(x)\label{diff int 3}
             \end{multline}
             Both, the l.h.s. and the r.h.s. represent a linear combination of conformal blocks. Two conformal blocks are linearly independent unless all of their dimensions (external and internal) coincide respectively. Thus, we can drop the integration and summation over $s_2$ in \eqref{diff int 3} to obtain
              
\begin{eqnarray}
            \boxed{ \hat{\co}_A^{s_1,s_2}(a_i,a)\cf_{aa'}  =\hat{\co}_B^{s_1,s_2}(a_i,a') \cf_{aa'}} \label{first order F}
             \end{eqnarray}

In the rest of the paper we study different consequences and look for solutions to equations \eqref{first order F}. One should stress, that these equations are by no means new. They represent  special (with some of dimensions set to degenerate values) cases of the pentagon identity required for consistency of the fusion algebra. In paper \cite{PT1} similar equations were also taken as a starting point for investigation of the fusion kernel. Solution was obtained there by establishing a connection  with quantum groups and identifying the fusion kernel with the Racah-Wiegner coefficients of a certain infinite-dimensional representation of $\mathcal{U}_q(sl_2)$ . Another approach suggested in \cite{Iorgov} builded on the relation between the fusion kernel and the connection coefficient of Painlev\'e VI. In the current work we take more direct attitude towards solution of \eqref{first order F} which hopefully brings additional clarity to the subject.

We should make one more remark about equations \eqref{first order F}. They originate from the pentagon identity which does not survive arbitrary renormalization of the fusion matrices. Non-trivial and important for our purposes fact is that matrices \eqref{Fa2} can be used in the pentagon identity. This can be verified directly by substituting \eqref{fmren},\eqref{norm} in the pentagon identity and observing complete cancellation of the normalization factors.

\section{Second order difference equation}
Equations \eqref{first order F} contain both shift operators in external and internal momenta. This makes them hard to deal with immediately. Instead, we can eliminate difference operators in external momenta at the cost of producing second-order difference operator in internal momentum. 

Apply operator $\hat{\co}_A^{-s_1,-s_2}(a_i,a)$ to both sides of equation \eqref{first order F}.  In the l.h.s. one obtains
\begin{multline}
\hat{\co}_A^{-s_1,-s_2}(a_i,a) \hat{\co}_A^{s_1,s_2}(a_i,a) \cf_{aa'} =\\\co_{A}^{-s_1,-s_2}(a_i,a)e^{\f{b}2(-s_1\p_{a_1}-s_2\p_{a_2})}\co_{A}^{s_1,s_2}(a_i,a)e^{\f{b}2(s_1\p_{a_1}+s_2\p_{a_2})}\cf_{aa'}=\\\co_{A}^{-s_1,-s_2}(a_i,a)\left(e^{\f{b}2(-s_1\p_{a_1}-s_2\p_{a_2})}\co_{A}^{s_1,s_2}(a_i,a)\right)\cf_{aa'}
\end{multline}
In the r.h.s. one obtains
\begin{multline}
\hat{\co}_A^{-s_1,-s_2}(a_i,a) \hat{\co}_B^{s_1,s_2}(a_i,a') \cf_{aa'}=\\\co_{A}^{-s_1,-s_2}(a_i,a)e^{\f{b}2(-s_1\p_{a_1}-s_2\p_{a_2})}\sum_s e^{-s\f{b}2\p_a'}\co_B^{s;s_1,s_2}(a_i,a')\cf_{aa'}=\\ \sum_s  e^{-s\f{b}2\p_a'}\left(e^{\f{b}2(-s_1\p_{a_1}-s_2\p_{a_2})}\co_B^{s;s_1,s_2}(a_i,a')\right)\hat{\co}_A^{-s_1,-s_2}\cf_{aa'}=\\\sum_s  e^{-s\f{b}2\p_a'}\left(e^{\f{b}2(-s_1\p_{a_1}-s_2\p_{a_2})}\co_B^{s;s_1,s_2}(a_i,a')\right)\hat{\co}_B^{-s_1,-s_2}\cf_{aa'}
\end{multline}
Now, in the l.h.s. there is no shift operator acting on $\cf_{aa'}$. In the r.h.s. we have two successive operators shifting internal momentum of $\cf_{aa'}$ by $\pm b/2$ each. Therefore, the r.h.s. can be decomposed into the sum of three terms each containing fusion kernel $\cf_{aa'}$ with $a'$ shifted by $-b,0$ or $+b$. Hence, equation
\begin{eqnarray}
\hat{\co}_A^{-s_1,-s_2}(a_i,a) \hat{\co}_A^{s_1,s_2}(a_i,a) \cm_{aa'}=\hat{\co}_A^{-s_1,-s_2}(a_i,a) \hat{\co}_B^{s_1,s_2}(a_i,a') \cf_{aa'}
\end{eqnarray}
can be converted into a second-order linear difference equation w.r.t. shifts in $a'$. Actually, there are four equations corresponding to each choice of signs for $s_1,s_2$. However, all these equations are equivalent to a single equation which we write in the following form
\begin{eqnarray}
 \boxed{\left(C_+(a,a')e^{b\p_{a'}}+C_0(a,a')+C_-(a,a')e^{-b\p_{a'}}\right)\cf_{aa'}=0\label{second order F C}} 
\end{eqnarray}
Functions $C_i(a,a')$ are straightforwardly calculable but still quite involved in a generic case. Explicit expressions can be found in appendix \ref{app func C}.

Analog of equation  \eqref{second order F C} was suggested in \cite{GMMnonpert}. There it was obtained from a different perspective. Namely, matrix model together with check-operator formalism. It was shown there, that the counterpart of equation \eqref{second order F C} for the toric fusion kernel is compatible with the result obtained in \cite{PT3}. We believe that the language of matrix models can be straightforwardly mapped to the one that we used so the coherence of the results in \cite{PT1} and \cite{GMMnonpert} is of no surprise: both the second order equation suggested in \cite{GMMnonpert} and the set of first order equations considered in \cite{PT1} are consequences of the pentagon identity. 
\section{Special cases}
\subsection{Ashkin-Teller case}
When the central charge is equal to one $c=1$ and all the external dimensions are equal to one sixteenth $\Delta_i=1/16$ conformal block, called Ashkin-Teller conformal block, is known in a closed form \cite{ZamAT}. This allows to calculate the corresponding fusion kernel explicitly. It appears to be the kernel of the Fourier transform
\begin{eqnarray}
\cf_{aa'}\begin{bmatrix}
1/16&1/16\\1/16&1/16
\end{bmatrix}\Big|_{c=1}=e^{-2\pi i aa'}\label{ATfk}
\end{eqnarray}
We can check equation \eqref{second order F C} against this special case. When $c=1$ and $\Delta_i=1/16$  functions $C_{i}(a,a')$ \eqref{functions C} reduce to
\begin{eqnarray}
C_{+}(a,a')=1,\quad C_0(a,a')=-2\cos{2\pi a}, \qquad C_-(a,a')=1
\end{eqnarray}
so that equation \eqref{second order F C} takes the following form ($b=1$ when $c=1$)
\begin{eqnarray}
\left( e^{\p_{a'}}-2\cos{2\pi a}+e^{-\p_{a'}}\right)\cf_{aa'}=0
\end{eqnarray}
Thus the Fourier kernel \eqref{ATfk} indeed solves this equation.
\subsection{Perturbative limit}
In papers \cite{GMMpert,Nemkov} the limit of large internal dimensions $a,a'\to \infty$ was studied in a first few perturbative orders of series expansion in powers of parameters $\frac1{a},\frac{1}{a'}$. It was found there, that in a suitable normalization the asymptotic form of the fusion kernel is again the Fourier kernel while the first several perturbative corrections are absent
\begin{eqnarray}
\cf_{aa'}\begin{bmatrix}
a_2&a_3\\a_1&a_4
\end{bmatrix}=e^{-2\pi i aa'+O(a^{-6})}\label{pertFK}
\end{eqnarray} 
We emphasize that \eqref{pertFK} is valid for any values of the external momenta $a_i$ and the central charge $c$.
In subsequent papers \cite{Lerda,GMMnonpert} it was proven that formula \eqref{pertFK} is correct in any order of perturbation theory. However, it is clear that the Fourier transform can not be the complete answer since the exacts results for the fusion kernel from \cite{PT1,PT2,Iorgov} clearly imply more complicated formula and thus existence of some non-perturbative corrections. The main goal of the present paper is to investigate these non-perturbative corrections.

Very important for the rest of the paper is the observation that functions $C_{i}(a,a')$ only depend on internal momenta $a,a'$ trough parameters $u ,u'$\footnote{For brevity, we denote functions of $u,u'$ by the same labels as functions of $a,a'$}
\begin{eqnarray}
\boxed{C_i(a,a')= C_i(u,u'),\quad u=e^{2\pi i b a},u'=e^{2\pi i b a'}}
\end{eqnarray}
This is a special feature of the chosen normalization.
  
 There is a region in the CFT parameter space where in the limit of $a,a'\to\infty$ parameters $u,u'\to0$ and are \emph{exponentially} small.  In the limit of vanishing $u'$ functions $C_i(u,u')$ become
 \begin{eqnarray}
 C_+(u,0)=1,\quad C_0(u,0)=-(u+u^{-1}e^{-2\pi i b^2})\equiv-2\cos{2\pi b a},\quad C_0(u,0)=e^{-2\pi i b^2}
 \end{eqnarray} 
 and equation \eqref{second order F C} multiplied by $e^{i\pi b^2}$ becomes
 \begin{eqnarray}
 \left(e^{i\pi b^2}e^{b\p_{a'}}-2\cos{2\pi b(a-b/2)}+e^{-i\pi b^2}e^{-b\p_{a'}}\right)\cf_{aa'}=0
 \end{eqnarray}
 Which is also solved by the Fourier kernel $\cf_{aa'}=e^{-2\pi i aa'}$.

 \section{Non-perturbative corrections}\label{nonpert}
 \subsection{The case of a general central change}
In the last section we saw that the perturbative Fourier-like form of the fusion kernel can be recovered in the limit $a,a'\to \infty$ while the small parameters effectively entering the equation are $u=e^{2\pi iba},u'=e^{2\pi iba'}$. It is our aim in this chapter to study the series expansion of the fusion kernel in powers of parameters $u,u'$. In fact, in order to approach \eqref{second order F C} iteratively it suffices to only work with expansion in one variable, say $u'$. Each $u'$-expansion term then will be an exact function of $u$.

Thus, we write
\begin{eqnarray}
\cf_{aa'}=e^{-2\pi i aa'}\sum_{ j}\phi^j (a,a')u'^{j} =e^{-2\pi i aa'}(\phi ^0(a,a')+u'\phi^1(a,a') +\dots)  
\end{eqnarray}
Factorizing the Fourier kernel out appears to be convenient. Here  $\phi^{j}(a,a') $ are yet arbitrary functions satisfying two requirements

$\bullet$  they do not grow too fast as $a,a'\to\infty$ so that $u'^{j}\phi_{j}(a,a')$ is indeed a small correction in this regime

$\bullet$ they do not non-trivially 'depend' on $u'$ so that any two corrections are indeed of a different order in $u'$.
 
Both these requirements are satisfied by polynomial in $a'$ functions $\phi_j(a,a')$. Explicit form of the fusion kernel in $c=1$ case also hints to the polynomial form of corrections.  Whether there are meaningful solutions that do not suit this ansatz is an important question which we do not address in the current paper. For the time being we consider only the polynomial ansatz.

We also introduce notation for the difference operator entering equation \eqref{second order F C}
\begin{eqnarray}
\widehat{C}(u,u')= C_{+}(u,u')e^{b\p_{a'}}+C_{0}(u,u')+C_{-}(u,u')e^{-b\p_{a'}}
\end{eqnarray}
and for its $u'$-expansion
\begin{eqnarray}
\widehat{C}(u,u')=\sum_ju'^j\widehat{C}^j(u)=\widehat{C}^0(u)+u'\widehat{C}^1(u)+\dots
\end{eqnarray}
Then, equation for the zeroth order of  \eqref{second order F C} reads
\begin{eqnarray}
 \wh{C}^{0}(u,u')\phi^0(a,a')=\left(u^{-1} e^{i\pi b^2}e^{b\p_{a'}}-(u^{-1}e^{i\pi b^2}+ue^{-i\pi b^2})+u e^{-i\pi b^2}e^{-b\p_{a'}}\right)\phi^{0}(a,a')=0 \label{zeroth order} 
\end{eqnarray}
 According to our ansatz, $\phi_0(a,a')$ is a polynomial in $a'$. One can easily check that difference operator $\widehat{C}^0$  lowers the degree of the polynomial on which it acts. This property makes $\widehat{C}^0$ similar to an ordinary differential operator and ensures that the only polynomial solution to \eqref{zeroth order} is constant in $a'$ function. In other words, $\phi^0(a,a')$ only depends on $a$
\begin{eqnarray}
\phi^0(a,a')=\phi^0(a)
\end{eqnarray} 
In its turn, the dependence on $a$ must be determined from the counterpart of equation \eqref{second order F C} containing shift operators in $a$.  

The first order correction satisfies
\begin{eqnarray}
\wh{C}^0(u) u' \phi^1(a,a')=-u'\wh{C}^1(u)\phi_0(a)\label{first order}
\end{eqnarray}
While $\wh{C}^0$ is relatively simple this is not the case already for $\wh{C}^1$. The explicit shape of $\wh{C}^1$ is not important for us at the moment, so we will not write it out. Instead, let us slightly rewrite  \eqref{first order}. 
\begin{align}
u'^{-1}\wh{C}^0(u)u'\phi^1(a,a')=-\wh{C}^1(u)\phi_0(a)\label{first order ref}
\end{align}
Consider operator
\begin{multline}
u'^{-1}\wh{C}^0 u' = u'^{-1}\left(u^{-1} e^{i\pi b^2}e^{b\p_{a'}}-(u^{-1}e^{i\pi b^2}+ue^{-i\pi b^2})+u e^{-i\pi b^2}e^{-b\p_{a'}}\right)u' =\\ \left(u e^{2i\pi b^2}e^{b\p_{a'}}-(u^{-1}e^{i\pi b^2}+ue^{-2i\pi b^2})+u^{-1} e^{-i\pi b^2}e^{-b\p_{a'}}\right) 
\end{multline} 
Note that this conjugated operator $\wh{C}^0\to u'^{-1}\wh{C}^0 u'$ no longer annihilates constant function of $a'$ and lowers the power of a polynomial. 
The r.h.s. of \eqref{first order ref} is independent of $a'$ and so must be the l.h.s. Any polynomial function $\phi(a,a')^{1}$ will maintain its degree if subjected to $u'^{-1}\wh{C}^0u'$. The only possibility left is constant in $a'$ function $\phi^1(a,a')$
\begin{eqnarray}
\phi^1(a,a')=\phi^1(a)
\end{eqnarray}
which is then unambiguously determined from \eqref{first order ref} to be
\begin{eqnarray}
\phi^1(a)=-\f{\wh{C}^1\phi_0(a)}{(u-u^{-1})(e^{i\pi b^2}-e^{-i\pi b^2})}\label{first order correction}
\end{eqnarray}
Our reasoning readily applies to higher corrections. Any function $\phi^{n}(a,a')$ in fact only depends on $a$ and can be straightforwardly determined if $\phi^{k<n}(a)$ are known. Moreover, since functions $\phi^{n}(a)$ are independent of $a'$ difference operators $\wh{C}^{k}$ acting on them can be represented simply as algebraic operators.

Let us also put it the other way. Assumption of $u'$-expansion and the polynomial ansatz appear to be so restrictive that the difference equation \eqref{second order F C} on the fusion kernel is reduced to an   algebraic equation on its $u'$-expansion coefficients
\begin{eqnarray}
\boxed{\cf_{aa'}=e^{-2\pi i aa'}\sum_{n}\phi^n(a)u'^n,\quad \sum_n \phi^{n}(a)C_n(u,u')u'^{n}=0}\label{non-unit c fk equations}
\end{eqnarray} 
where we denoted
\begin{empheq}[box=\fbox]{multline}
C_n(u,u')=u'^{-n}e^{2\pi i aa'}\wh{C}(u,u')u'^{n}e^{-2\pi i aa'}\circ 1=\\e^{2\pi i b^2 n}u^{-1}C_+(u,u')+C_0(u,u')+e^{-2\pi i b^2 n}uC_-(u,u')
\end{empheq}

Thus we propose explicit and simple way of finding non-perturbative corrections to the general fusion kernel. Unfortunately, we can not test these against the results of \cite{PT1,PT2} since it is not clear for us how to perform a series expansion of the answer there.
Luckily, in the $c=1$ case to be discussed in the next section explicit comparison with independent results from \cite{Iorgov} is possible. 
\subsection{The case of the unit central charge}
The case of the unit central charge, corresponding to $b=1$ in our notation \eqref{central charge}, requires separate consideration. One can see that the first order correction \eqref{first order correction} to the fusion kernel is singular in the limit $b\to1$. 

The reason is that at the point $b=1$ parameters $u=e^{2\pi i b a},u'=e^{2\pi i b a'}$ become \textit{periodic} functions of $a,a'$ with the \textit{unit} period\footnote{In the sequel, we sometimes call these simply 'periodic functions', omitting specification of the period}. Thus, they are unaffected by the shift operators in \eqref{second order F C}. Consequently, when $b=1$ operator $u'^{-n}\wh{C}^0u'^{n}=\wh{C}^0$ again degenerates to a 'differential' operator, i.e. lowers the degree of a polynomial. This forces us to rethink equation \eqref{first order ref} which can now be casted simply as
\begin{eqnarray}
\wh{C}^0(u)\phi^1(a,a')=-\wh{C}^1(u)\phi^0(a)\label{first order b=1}
\end{eqnarray}
Independent of $a'$ function $\phi^1(a,a')$ is no longer a solution to \eqref{first order b=1} since a constant in $a'$ function is now annihilated by $\wh{C}^0$. Instead, in order to give out a zeroth order polynomial operator $\wh{C}^0$ must take in a linear function
\begin{eqnarray}
\phi^1(a,a')=\phi^{10}(a)+a'\phi^{11}(a)
\end{eqnarray}
function $\phi^{11}(a)$ can be straightforwardly determined from \eqref{first order b=1} to be
\begin{eqnarray}
\phi^{11}(a)=\f{\wh{C}^1(u)\phi^0(a)}{u-u^{-1}}
\end{eqnarray}
while function $\phi^{01}(a)$ can not be determined from equations \eqref{second order F C},\eqref{first order b=1}.

Equation for the second order correction is
\begin{eqnarray}
\wh{C}^0(u)\phi^2(a,a')=-\wh{C}^1(u)\phi^1(a,a')-\wh{C}^2(u)\phi^0(a,a')\label{second order b=1}
\end{eqnarray}
and now we have a linear in $a'$ function in the r.h.s. so $\phi^2(a,a')$ must be quadratic
\begin{eqnarray}
\phi^2(a,a')=\phi^{20}(a)+a'\phi^{21}(a)+a'^2\phi^{22}(a)
\end{eqnarray}
Functions $\phi^{21}(a),\phi^{22}(a)$ can be determined from \eqref{second order b=1} while $\phi^{20}(a)$ can not.

Hereby, we arrive to the following form of the fusion kernel in the $b=1$ case
\begin{eqnarray}
\cf_{aa'}=e^{-2\pi i aa'}\sum_n \phi^n(a,a')u'^n, \quad \phi^n(a,a')=\sum_{k=0}^{n}a'^k\phi^{nk}(a)\label{ansatz b=1}
\end{eqnarray}
where each function $\phi^{n\geq1,k\geq1}(a)$ can be determined recursively if $\phi^{n'<n,k'<k}(a)$ are known. Functions $\phi^{n0}(a)$ remain unfixed for any $n$ for a simple reason. If $b=1$ they enter corrections which are periodic functions of $a'$ and therefore remain unnoticeable by equation \eqref{second order F C}.

One important observation remains to be made here. While the $c=1$ fusion kernel itself contains polynomials of arbitrary degree in $a'$ its logarithm is a 'linear' function of $a'$\footnote{In the sequel we often refer to such functions simply as 'linear'.}
\begin{eqnarray}
\log{\cf_{aa'}}=A(a,a')+a'B(a,a')\label{linear log}
\end{eqnarray}
Here $A(a,a')$ and $B(a,a')$ are some periodic functions of $a'$ with the unit period. Statement \eqref{linear log} does not follow from our previous considerations. One can directly check it in the several first orders of the $u'$-expansion. In the next section we give a simple explanation of this phenomenon.
\section{Exact solution for the $
c=1$ fusion kernel}
\subsection{Dependence on internal dimensions from second-order equation}
As we have seen in the previous section, the logarithm of the fusion kernel appears to be a 'linear' (we regard to periodic functions of $a'$ as constants in this context) function of $a'$. Fusion kernel should be essentially symmetric function w.r.t. exchange $a\leftrightarrow a'$ and thus we can safely assume that the dependence on $a$ is also linear. Moreover, one can notice that in the first-order equations \eqref{first order F} external and internal momenta also enter rather symmetrically so all our arguments readily apply to the dependence on the external momenta.

Hereby, we propose the following ansatz for $c=1$ fusion kernel
\begin{eqnarray}
\boxed{\cf_{aa'}=\exp\left(-2\pi i aa'+aP +a'P' +\sum_{i=1}^{4}a_iP_i +R\right)\label{linear ansatz}}
\end{eqnarray}
where $P,P',P_i,R$ all are functions periodic w.r.t. any external $a_i$ or internal $a,a'$ momenta (with the unit period).

There is a short way to come up with the ansatz \eqref{linear ansatz}. We can note the in the case of $b=1$ all coefficients in equations \eqref{first order F} are periodic functions with the unit period. So we are effectively solving  difference equations with \textit{constant} coefficients. It is then straightforward to motivate \eqref{linear ansatz}.
 
Now, upon substitution of \eqref{linear ansatz} equation \eqref{second order F C} reduces to a quadratic equation w.r.t. $e^{P'}$
\begin{eqnarray}
u^{-1} C_+(u,u')e^{P' }+C_0(u,u')+u C_-(u,u')e^{-P' }=0
\end{eqnarray}
which have two solutions
\begin{eqnarray}\boxed{
P'=\log\left\{\f{-C_0(u,u')\pm\s{C_0(u,u')^2-4C_+(u,u')C_-(u,u')}}{2C_+(u,u')}\right\}\label{P'}+\log{u}}
\end{eqnarray}
With the help of explicit expressions \eqref{functions C} it can be shown that only one branch of the square root here is actually a periodic function of $a$ (the $\log{u}$ term gets canceled) while the other is not. 

Function $P$ should be determined from the counterpart of equation \eqref{second order F C} with shift operators in $a$ instead of $a'$.   
\subsection{Dependence on external dimensions from first-order equations}
Functions $P_i$ can not be determined from the second order difference equation with shift operators in internal momenta. However, we can find them by returning back to the first-order equations containing shifts in external momenta \eqref{first order F}. The only complication in extracting them is due to \eqref{first order F} containing simultaneous shifts in different external momenta. We can perform the same procedure that allowed us to derive second-order equation \eqref{second order F C}.  Consider the following expression
\begin{eqnarray}
\hat{\co}_A^{-s_1,s_2}(a_i,a)\hat{\co}_A^{s_1,s_2}(a_i,a)\cm_{aa'}=\hat{\co}_A^{-s_1,s_2}(a_i,a)\hat{\co}_B^{s_1,s_2}(a_i,a')\cm_{aa'}   \label{two A op}
\end{eqnarray}
Recall that indices $s_1,s_2$ correspond to shifts in momenta $a_1$ and $a_2$ respectively. Hence, in the l.h.s. of \eqref{two A op} shifts in $a_1$ in the fusion kernel cancel out leaving only the shift in $a_2$. In the r.h.s. we can convert the action of the $A$-operator to the action of the $B$-operator so that only shift operators in $a'$ enter the r.h.s. In analogy with \eqref{second order F C} we can write the result as
\begin{eqnarray}
e^{\p_{a_2}}\cf_{aa'}=\left(E_+(u,u')e^{b\p_{a'}}+E_0(u,u')+E_-(u,u')e^{-b\p_{a'}}\right)\cf_{aa'}
\end{eqnarray}
Functions $E_i$ can be found in appendix \ref{app func E}

Since function $P'$ is already found  we can compute the r.h.s. explicitly and thus immediately obtain
\begin{eqnarray}
\boxed{P_2=\log{\left(E_+(u,u')e^{P}+E_0(u,u')+E_-(u,u')e^{-P}\right)}}\label{P2 solution}
\end{eqnarray}
Computing functions $P_1,P_3,P_4$ is a straightforward procedure going thorough the same steps that we did for $P_2$. Function $R$ can not be determined from our considerations and represents unavoidable ambiguity inherent in solutions to difference equations.

Hereby, with the help of ansatz \eqref{linear ansatz} we were able to exactly solve equations \eqref{first order F} in the case of the unit central charge. In the next section we present the formula for $c=1$ fusion kernel suggested in \cite{Iorgov}. After that, we compare these two answers to find a complete agreement.
\subsection{Other exactly solvable cases}
For any $b\in \mathbb{N}$ the value of shifts in difference equations \eqref{first order F} is an integer multiple of the coefficients period.  This allows to find exactly the same formula for the fusion kernel as for $b=1$ case considered above. The only difference is that we now must substitute $b=n\in\mathbb{N}$ instead of $b=1$ in \eqref{functions C},\eqref{functions E}.
\section{$c=1$ fusion kernel and the theory of Painlev\'e VI}
\subsection{Relation between the connection coefficient and the fusion kernel}
In paper \cite{Iorgov} formula for the generic $c=1$ fusion kernel was suggested. Here we intend to give a brief review of the approach and the results obtained in \cite{Iorgov}.

The central role is played by the relation between $c=1$ conformal blocks and $\tau$ function of Painlev\'e VI. Namely, conformal block $B_a\left[\begin{array}{cc}a_2&a_3\\a_1&a_4\end{array};x\right]$ appears to be a Fourier series coefficient of the $\tau$ function expansion w.r.t. to one of its parameters
\begin{eqnarray}
\tau=\chi_0(\vec{\theta},\sigma_{0t},\sigma_{1t})\sum_{n\in \mathbb{Z}}C\left[\begin{array}{cc}\theta_t&\theta_1\\\theta_0&\theta_\infty\end{array};\sigma_{0t}+n\right]s_{0t}^n B_{\sigma_{0t}+n}\left[\begin{array}{cc}\theta_t&\theta_1\\\theta_0&\theta_\infty\end{array};t\right]=\nn\\
\chi_1(\vec{\theta},\sigma_{0t},\sigma_{1t})\sum_{n\in \mathbb{Z}}C\left[\begin{array}{cc}\theta_t&\theta_0\\\theta_1&\theta_\infty\end{array};\sigma_{1t}+n\right]s_{1t}^n B_{\sigma_{1t}+n}\left[\begin{array}{cc}\theta_t&\theta_0\\\theta_1&\theta_\infty\end{array};1-t\right]\label{tau expansion}
\end{eqnarray}
All notation appeared here will be explained below. Note that in the second line instead of original $s$-channel conformal blocks there are $t$-channel conformal blocks. Since a standard conformal block is a power series in $t$, expression in the first line can be used to obtain the expansion of the $\tau$ function near point $t=0$ while the second line produces expansion near $t=1$. Relation between these expansions is captured by the quantity
\begin{equation}
\chi_{01}=\chi_0^{-1}\chi_1
\end{equation}
which is called the connection coefficient.

It also convenient to use renormalized $\chi_{01}$
\begin{eqnarray}
\bar{\chi}_{01}=\chi_{01}\f{C\left[\begin{array}{cc}\theta_t&\theta_0\\\theta_1&\theta_\infty\end{array};\sigma_{1t}\right]}{C\left[\begin{array}{cc}\theta_t&\theta_1\\\theta_0&\theta_\infty\end{array};\sigma_{0t}\right]}
\end{eqnarray}

 Before going into deciphering of \eqref{tau expansion} let us spell out the exact relation between connection coefficient and the $c=1$ fusion kernel
\begin{equation}
\boxed{
\cf_{aa'}\begin{bmatrix}a_2&a_3\\a_1&a_4\end{bmatrix}=\bar{\chi}_{01}(\vec{\theta},\sigma_{0t},\sigma_{1t})\mu(\vec{\theta},\sigma_{0t},\sigma_{1t})\label{fktocc}}
\end{equation}
Here $\chi_{01}$ is the main object of pursuit in \cite{Iorgov},  $\mu$ is some relatively simple combination of trigonometric functions and $C$ are structure constants expressed in terms of Barnes double gamma function. Explicit definitions of all these functions and relation between parameters used in the l.h.s. and r.h.s. will be given below.
\subsection{Difference equation for the connection coefficient}
By definition, connection coefficient is a ratio of two Fourier-expansions. It is straightforward to see from \eqref{tau expansion} that $\chi_{01}$ satisfies the following difference equations
\begin{eqnarray}
\chi_{01}(\vec{\theta},\sigma_{0t}+1,\sigma_{1t})=s_{0t}^{-1}\chi_{01}(\vec{\theta},\sigma_{0t},\sigma_{1t})\nn\\ 
\chi_{01}(\vec{\theta},\sigma_{0t},\sigma_{1t}+1)=s_{1t} \chi_{01}(\vec{\theta},\sigma_{0t},\sigma_{1t})\label{diff eq fo}
\end{eqnarray} 
 
\subsection{Formula for the connection coefficient}
It was suggested in \cite{Iorgov} that under certain assumptions the unique solution to equations \eqref{diff eq fo} is given by the function 
\begin{align} 
\bar{\chi}_{01}(\vec{\theta},\sigma_{0t},\sigma_{1t})=&\prod_{k=1}^{4} \f{\hat{G}(\omega_++\nu_k)}{\hat{G}(\omega_++\lambda_k)}\times\nn\\&\prod_{\e,\e'=\pm}\frac{G(1+\e \sigma_{1t}+\e'\theta_t-\e\e'\theta_1)G(1+\e \sigma_{1t}+\e'\theta_0-\e\e'\theta_\infty)}{G(1+\e \sigma_{0t}+\e'\theta_t+\e\e'\theta_0)G(1+\e \sigma_{0t}+\e'\theta_1+\e\e'\theta_\infty)} \prod_{\e=\pm}\f{G(1+2\e \sigma_{0t})}{G(1+2\e \sigma_{1t})}\label{chi01}
\end{align}
where $G(x)$ is the Barnes double gamma function satisfying
\begin{eqnarray}
G(x+1)=G(x)\Gamma(x)
\end{eqnarray}
 and $\hat{G}(x)$ stands for
 
 \begin{eqnarray}
 \hat{G}(x)=\frac{G(1+x)}{G(1-x)}
 \end{eqnarray}
In expression \eqref{chi01}  the factor in the first line is the most essential part of the result. The factors in the second line can be removed by the appropriate renormalization of conformal blocks.  
 
These statements are all we need so far. In order to make use of them it suffices to untangle all the auxiliary notation used.
\subsection{Notation I}  
 
Parameters $\sigma_{0t},\sigma_{1t}$ and $\theta_0,\theta_t,\theta_1,\theta_\infty$ denote the same quantities as we used for internal and external momenta while $t$ is the four-point cross ratio
\begin{align}
\{\theta_0,\theta_t,\theta_1,\theta_\infty\}&\sim  \{a_1,a_2,a_3,a_4\}\nn\\
\{\sigma_{0t},\sigma_{1t}\}&\sim  \{a,a'\}\nn\\
t&\sim x
\end{align}
Next, the structure constants are given by
\begin{eqnarray}
 C\left[\begin{array}{cc}\theta_t&\theta_1\\\theta_0&\theta_\infty\end{array};\sigma\right]=\f{\prod_{\e,\e'=\pm}G(1+\theta_t+\e \theta_0+\e'\sigma)G(1+\theta_1+\e \theta_\infty+\e'\sigma)}{\prod_{\e=\pm}G(1+2\e\sigma)}
\end{eqnarray}

It remains to explain what are parameters $s_{0t},s_{1t}$ in \eqref{diff eq fo} and what are $\omega_+,\nu_k,\lambda_k$ in \eqref{chi01}. In order to efficiently do that we will have to introduce another set of auxiliary notation.
\subsection{Notation II}
Introduce 
\begin{eqnarray}
p_\mu=2\cos{2\pi\theta_\mu},\quad \mu=0,t,1,\infty\nn\\
p_{\mu\nu}=2\cos{2\pi \sigma_{\mu\nu}}, \quad \mu\nu=0t, 1t, 01
\end{eqnarray}
Here new quantity $\sigma_{01}$ is not independent of $\sigma_{0t},\sigma_{1t}$ but related to them via quadratic  equation $W=0$ with
\begin{eqnarray}
W=p_{0t}p_{1t}p_{01}+p_{0t}^2+p_{1t}^2+p_{01}^2-\omega_{0t} p_{0t}-\omega_{1t} p_{1t}-\omega_{01} p_{01}+\omega_4-4
\end{eqnarray}
and $\omega$'s are given purely in terms of $p_\mu$/$\theta_\mu$
\begin{eqnarray}
\omega_{0t}=p_{0}p_t+p_1 p_\infty\nn\\
\omega_{1t}=p_1p_{t}+p_0 p_\infty\nn\\
\omega_{01}=p_{0}p_1+p_t p_\infty\nn\\
\omega_4=\prod_{\mu}p_\mu+\sum_\mu p_\mu^2
\end{eqnarray}
Then, functions $s_{0t},s_{1t}$ are defined by
\begin{eqnarray}
s_{0t}^{\pm}=\f{q_{01}e^{\pm2\pi i\sigma_{0t}}-q_{1t}}{16\prod_{\e=\pm}\sin(\theta_t\mp\sigma_{0t}+\e\theta_0)\sin(\theta_1\mp\sigma_{0t}+\e\theta_\infty)}\nn\\
s_{1t}^{\pm}=\f{q_{01}e^{\mp2\pi i\sigma_{1t}}-q_{0t}}{16\prod_{\e=\pm}\sin(\theta_t\mp\sigma_{1t}+\e\theta_1)\sin(\theta_0\mp\sigma_{1t}+\e\theta_\infty)}
\end{eqnarray}
here $q_{\mu\nu}=\f{\p W}{\p p_{\mu\nu}}$
\begin{eqnarray}
q_{01}=2p_{01}+p_{0t}p_{1t}-\omega_{01}\nn\\
q_{0t}=2p_{0t}+p_{01}p_{1t}-\omega_{0t}\nn\\
q_{1t}=2p_{1t}+p_{0t}p_{01}-\omega_{1t}
\end{eqnarray}
\subsection{Notation III}
Parameters $\nu_k$ and $\lambda_k$ are  given as a simple linear functions of $\vec{\theta}$ and $\sigma_{0t},\sigma_{1t}$
\begin{align}
&\nu_1=\sigma_{0t}+\theta_0+\theta_t,\quad &\lambda_1&=\theta_0+\theta_t+\theta_1+\theta_\infty\nn\\
&\nu_2=\sigma_{0t}+\theta_1+\theta_\infty,\quad &\lambda_2&=\sigma_{0t}+\sigma_{1t}+\theta_0 +\theta_1 \nn\\
&\nu_3=\sigma_{1t}+\theta_0+\theta_\infty,\quad &\lambda_3&=\sigma_{0t}+\sigma_{1t}+\theta_t +\theta_\infty\nn\\
&\nu_4=\sigma_{1t}+\theta_t+\theta_1,\quad &\lambda_4&=0
\end{align}
The thing to note here is that $\sum_{k}\nu_k=\sum_{k}\lambda_k$ and this quantity deserved a separate notation $\nu_\Sigma=\f12\sum_{k}\nu_k$.

Parameter $\omega_+$ depends on $\nu,\lambda$ and is defined as follows. Consider equation
\begin{equation}
\prod_{k}(1-ze^{2\pi i \nu_k})=\prod_{k}(1-ze^{2\pi i \lambda_k})\label{cubic}
\end{equation}
Since $\nu_\Sigma=\lambda_\Sigma$ this is a cubic equation w.r.t. to $z$. It admits one trivial solution $z=0$ and two non-trivial solutions $z_{\pm}$. Parameters $\omega_{\pm}$ are defined by
\begin{equation}
z_{\pm}=e^{2\pi i \omega_{\pm}}\label{zpm}
\end{equation}
Non-trivial solutions $z_\pm$ can be expressed in the following form
\begin{eqnarray}
z_\pm=\f{4\sin{2\pi \sigma_{0t}}\sin{2\pi \sigma_{1t}}+4\sin{2\pi\theta_t}\sin{2\pi\theta_\infty}+4\sin{2\pi\theta_0}\sin{2\pi\theta_1}\pm q_{01}}{2\sum_{k}\left(e^{2\pi i(\nu_\Sigma-\nu_k)}-e^{2\pi i(\nu_\Sigma-\lambda_k)}\right)}
\end{eqnarray} 
Finally, function $\mu$ appearing in \eqref{fktocc} reads
\begin{equation}
\mu(\vec{\theta},\sigma_{0t},\sigma_{1t})=-\f{4\sin{2\pi\sigma_{0t}}\sin{2\pi\sigma_{1t}}}{q_{01}}
\end{equation}
This completes the definition of formulas \eqref{chi01},\eqref{fktocc} up to the ambiguity to be discussed right below.

 One should make several more remarks. The r.h.s. of \eqref{chi01} is periodic function of $\omega_+$ which enables to use arbitrary branch of solution to \eqref{zpm}. However, one overall uncertainty remains. We have introduced $p_{01}$ as a solution to a quadratic equation. So there are actually two different possibilities for $p_{01}$. If one of them is chosen then parameters $q_{01}$ and $z_+$ are defined unambiguously.

 Both choices of $p_{01}$ provide the connection coefficient $\chi_{01}$ that solves \eqref{diff eq fo} but only one of them is relevant for the fusion kernel. Namely, in the formula \eqref{fktocc} only such $p_{01}$ must be chosen for which $s_{0t}\to0$ as $\Im(\sigma_{1t})\to\infty$. This situation is reminiscent of the doubling of solutions that we had in \eqref{P'}.
 \section{Comparison of the formulas for $c=1$ fusion kernel}
 \subsection{Series expansions}
  Formula for the $c=1$  fusion kernel  \eqref{fktocc} is amenable to  explicit expansion in powers of the 'non-perturbative' parameters $u,u'$. When comparing two fusion kernels one must take in account possible difference in normalizations
\begin{eqnarray}
\cf_{aa'}\begin{bmatrix}
a_2&a_3\\a_1&a_4
\end{bmatrix}\sim N_a\begin{bmatrix}
a_2&a_3\\a_1&a_4
\end{bmatrix}\cf_{aa'}\begin{bmatrix}
a_2&a_3\\a_1&a_4
\end{bmatrix}N^{-1}_{a'}\begin{bmatrix}
a_2&a_1\\a_3&a_4
\end{bmatrix}
\end{eqnarray} In order to get rid of this ambiguity we only consider the diagonal elements $a=a'$ and set $a_1=a_3$ so that normalization factors  cancel out. Furthermore, for the sake of simplicity and visibility we additionally specify external momenta to be $a_1=a_3=a_4=1/4$ only leaving $a_2$ as a free parameter.
It is also more convenient technically  to deal with logarithmic derivative of the fusion kernel rather then with the kernel itself.

With these assumptions, several first terms in $u$-expansion of the $c=1$ fusion kernel from \cite{Iorgov} are
\begin{multline}
\p_a \log{\cf_{aa}}\begin{bmatrix}
a_2&1/4\\1/4&1/4
\end{bmatrix}=4 i a \pi +2 \pi  \left(\cos \left(2 \pi  a_2\right)+4 i a \left(-1+\sin
   \left(2 \pi  a_2\right)\right)+4 \cos \left(2 \pi  a_2\right)
   a_2\right) u-\\4 \left(\pi  \cos \left(2 \pi  a_2\right) \left(1-4 i a
   \cos \left(2 \pi  a_2\right)+\sin \left(2 \pi  a_2\right)+4 \sin
   \left(2 \pi  a_2\right) a_2\right)\right) u^2+O\left(u^3\right)
\end{multline}
Series expansion of the expression suggested for $c=1$ fusion kernel in the current work can of course be calculated both iteratively as described in section \ref{nonpert} and directly from \eqref{linear ansatz} with $P$ and $P_2$ given by \eqref{P'},\eqref{P2 solution}\footnote{One should also take in account that when $a'=a$ and $a_1=a_3$ one has $P(a,a')=P'(a,a')$}. The result is
\begin{multline}
\p_a \log{\cf_{aa}}\begin{bmatrix}
a_2&1/4\\1/4&1/4
\end{bmatrix}=4 i a \pi +\left(4 (1+2 i a \pi ) \left(-1+\sin \left(2 \pi 
   a_2\right)\right)+8 \pi  \cos \left(2 \pi  a_2\right) a_2\right) u\nn\\+4
   \cos \left(2 \pi  a_2\right) \left((1+4 i a \pi ) \cos \left(2 \pi 
   a_2\right)-4 \pi  \sin \left(2 \pi  a_2\right) a_2\right)
   u^2+O\left(u^3\right)
\end{multline}
One can see that these two expressions exactly coincide modulo terms periodic with respect to $a$ and $a_2$. Such terms were not fixed by our considerations. We checked such expansions in a few more orders and found no discrepancies.
\subsection{Closer look at the Painlev\'e connection coefficient}
Denote first factor in \eqref{chi01} by $\cm$ (and recall that the rest can be removed by renormalization)
\begin{eqnarray}
\cm=\prod_{k=1}^{4} \f{\hat{G}(\omega_++\nu_k)}{\hat{G}(\omega_++\lambda_k)}\label{core}
\end{eqnarray}
There is a reflection formula for the Barnes gamma function
\begin{eqnarray}
\log{\hat{G}(x)}=\log{\frac{G(1+x)}{G(1-x)}}=x\log2\pi-\int_0^x\pi y\cot\pi y\label{reflection}
\end{eqnarray}
Thus
\begin{eqnarray}
\p_x\log{\hat{G}(x)}=\log2\pi-\pi x\cot\pi x\label{dreflection}
\end{eqnarray}

Using \eqref{reflection} it is easy to see that the logarithmic derivative of \eqref{core} only contains the elementary functions. It can be explicitly demonstrated that this derivative is 'linear' in $a,a',a_i$ function with periodic coefficients of the unit period, as in \eqref{linear log}. Our formula \eqref{linear ansatz} imply that the 'linear' monomial in this expression after integration can be casted purely in terms of the elementary functions. Indeed, an alternative representation in the spirit of \eqref{core} reads 
\begin{eqnarray}
\cm = \prod_{k=1}^{4} \frac{\left(\sin\pi(\omega_++\nu_k)\right)^{-\nu_k}}{\left(\sin\pi(\omega_++\lambda_k)\right)^{-\lambda_k}}\label{spec} \times \text{periodic}\label{per_simple}
\end{eqnarray}
The remaining periodic factor probably can not be simplified in a similar way. Derivation of \eqref{per_simple} can be found in appendix \ref{app_simplifying}

\section{Summary and discussion}

In the present paper we made an attempt to gain more quantitative understanding of the fusion algebra for non-degenerate conformal blocks in Liouville theory. We have identified quantities $u=e^{2\pi i b a}, u'=e^{2\pi i b a'}$ as 'non-perturbative' parameters controlling large $a,a'$ expansion and showed that in the limit of vanishing $u,u'$ one recovers the perturbative Fourier-like shape of the fusion kernel.

 For the case of a non-unit central charge we proposed simple algebraic equations \eqref{non-unit c fk equations} together with a certain ansatz which allows one to compute non-perturbative corrections. For the case of the unit central charge we suggested the following formula as an exact expression
\begin{align}
\boxed{\cf_{aa'}\begin{bmatrix}
a_2&a_3\\a_1&a_4
\end{bmatrix}=\exp\left(-2\pi i aa'+aP +a'P' +\sum_{i=1}^{4}a_iP_i +R\right)}\label{ff}
\end{align}
Here $P,P',P_i$ are cumbersome but straightforward functions  which can be expressed in terms of the elementary functions (for example see formulas \eqref{P'},\eqref{functions C} for $P'$). Function $R$ is periodic with the unit period in all external and internal momenta. This function remains unfixed by our considerations. With some restrictions imposed, we have checked this formula against the one derived in \cite{Iorgov} and found a complete agreement. This looks rather non-trivial since formula \eqref{chi01} pretends to posses a much more complicated structure. However, we showed that the periodic part of this formula after appropriate renormalization can be restated in terms of the elementary functions \eqref{spec}. 

Formula \eqref{ff} not only describes the $c=1$ case, but is also valid for any central charge corresponding to $b\in\mathbb{N}$. One only needs to use the appropriate values of $b$ in \eqref{functions C},\eqref{functions E}.

In the current paper we did not address a lot of important and interesting questions.  First group of them concerns more thorough investigation of the $c=1$ formula for the fusion kernel. It would be interesting to explicitly compare \eqref{linear ansatz} and \eqref{chi01}, including normalization factors and the periodic remainder. Second group of questions is related to the extension of our analysis to $b\notin \mathbb{N}$ case. With the help of equations \eqref{first order F} it is possible to study various expansions of the fusion kernel, not only in powers of $u,u'$. In the limit $b\to0$ corrections are described by differential equations. In the limit $b\to1$ corrections satisfy difference equations with periodic coefficients which are exactly solvable. Working out all these cases should yield a better understanding of the non-degenerate fusion algebra. And of course, learning to obtain all these expansions from the exact result of \cite{PT1,PT2} is highly desirable.

\section*{Acknowledgements}
The author is grateful to Alexei Morozov, Andrei Mironov, and Dmitry Galakhov for their great assistance. This work is partly supported by the grant NSh - 1500.2014.2, by RFBR 13-02-00457, and by RFBR 14-02-31372 mol\_a. 
 
\appendix
\section{Functional coefficients in difference equations}
\subsection{Functions $C_+$, $C_0$, $C_-$} \label{app func C}
We can divide all functions $C_{i}(a,a')$ by a common factor which corresponds to an overall normalization of equation \eqref{second order F C}. If we chose
\begin{eqnarray}
C_{+}=1
\end{eqnarray}
then the other two functions are given by
\begin{align}
C_0(u,u')&=-\frac{e^{i b \pi  \left(a_1+a_2-a_3-a_4\right)} u' \left(\frac{e^{-i b^2 \pi }}{u'}-e^{i b^2 \pi } u'\right)}{\left(-1+e^{2 i b \pi 
   \left(a_2-a_3\right)} u'\right) \left(-1+e^{2 i b \pi  \left(a_1-a_4\right)} u'\right)}\times\nn\\&\left(\frac{e^{-i b \pi  \left(b-a_1-a_2-a_3-a_4\right)} \left(-1+e^{2 i b \pi  \left(b-a_2-a_3\right)} u'\right) \left(-1+e^{2 i b \pi 
      \left(b-a_1-a_4\right)} u'\right)}{e^{2 i b^2 \pi }-\left(u'\right)^2}+\right.\nn\\&\left.\frac{e^{-i b \pi  \left(3 b-3 a_1-3 a_2+a_3+a_4\right)} \left(e^{2 i b \pi  \left(2 b-a_1-a_2\right)}-u\right) \left(-1+e^{2 i b \pi 
         \left(b-a_1-a_2\right)} u\right) \left(-1+\left(u'\right)^2\right)}{u \left(e^{2 i b \pi  \left(a_2-a_3\right)}-u'\right) \left(e^{2 i b \pi
          \left(a_1-a_4\right)}-u'\right)}+\right.\nn\\&\left.\frac{e^{-i b \pi  \left(3 b-3 a_1-3 a_2+a_3+a_4\right)} \left(e^{2 i b \pi  \left(2 b-a_2-a_3\right)}-u'\right) \left(-e^{2 i b \pi 
             \left(b-a_2+a_3\right)}+u'\right)}{\left(-e^{2 i b \pi  \left(a_2-a_3\right)}+u'\right) \left(-e^{2 i b \pi 
             \left(a_1-a_4\right)}+u'\right)}\times\right.\nn\\&\left.\frac{\left(-e^{2 i b \pi  \left(2 b-a_1-a_4\right)}+u'\right) \left(-e^{2 i b \pi  \left(b-a_1+a_4\right)}+u'\right)
                \left(-1+\left(u'\right)^2\right)}{\left(-e^{2 i b^2 \pi }+\left(u'\right)^2\right) \left(-e^{4 i b^2 \pi }+\left(u'\right)^2\right)} \right)\nn\\
                C_-(u,u')&=\frac{e^{2 i b \pi  \left(a_1+a_2-a_3-a_4\right)} \left(e^{2 i b \pi  \left(2 b-a_2-a_3\right)}-u'\right) \left(e^{2 i b \pi  \left(2
                   b-a_1-a_4\right)}-u'\right)}{\left(e^{2 i b \pi  \left(a_2-a_3\right)}-u'\right) \left(e^{2 i b \pi  \left(a_1-a_4\right)}-u'\right)
                   \left(-1+e^{2 i b \pi  \left(a_2-a_3\right)} u'\right)}\times\nn\\&\frac{\left(-e^{2 i b \pi  \left(a_2+a_3\right)}+u'\right) \left(-e^{2 i b \pi  \left(a_1+a_4\right)}+u'\right)
                      \left(-1+\left(u'\right)^2\right) \left(-1+e^{2 i b^2 \pi } \left(u'\right)^2\right)}{\left(-1+e^{2 i b \pi  \left(a_1-a_4\right)} u'\right)
                      \left(e^{4 i b^2 \pi }-\left(u'\right)^2\right) \left(e^{6 i b^2 \pi }-\left(u'\right)^2\right)}\label{functions C}
\end{align}
where $u=e^{2\pi i ba}, u'=e^{2\pi i b a'}$
\subsection{Functions $E_+$, $E_0$, $E_-$}\label{app func E}
functions $E_i$ entering equation \eqref{P2 solution} are
\begin{align}
E_+(u,u')&=\frac{e^{-2 i \pi  \left(a_2-a_3-a_4\right)} u^2 \left(e^{2 i \pi 
   \left(a_2-a_3\right)}-u'\right){}^2 \left(e^{2 i \pi 
   \left(a_1-a_4\right)}-u'\right) \left(-1+e^{2 i \pi 
   \left(a_1-a_4\right)} u'\right)}{\left(-1+e^{2 i \pi 
   \left(a_1-a_2\right)} u\right) \left(-1+e^{2 i \pi 
   \left(a_1+a_2\right)} u\right) \left(-1+\left(u'\right)^2\right)^2}\nn\\
   E_0(u,u')&=-\frac{2 e^{2 i \pi  \left(a_1-a_2\right)} u \left(e^{2 i \pi 
      \left(a_2-a_3\right)}-u'\right) \left(-1+e^{2 i \pi 
      \left(a_2+a_3\right)} u'\right) \left(-2 \cos \left(2 \pi  a_4\right)
      u'+\cos \left(2 \pi  a_1\right)
      \left(1+\left(u'\right)^2\right)\right)}{\left(-1+e^{2 i \pi 
      \left(a_1-a_2\right)} u\right) \left(-1+e^{2 i \pi 
      \left(a_1+a_2\right)} u\right) \left(-1+\left(u'\right)^2\right)^2}\nn\\
   E_-(u,u')&=\frac{e^{-2 i \pi  \left(a_2+a_3+a_4\right)} \left(e^{2 i \pi 
      \left(a_1+a_4\right)}-u'\right) \left(-1+e^{2 i \pi 
      \left(a_2+a_3\right)} u'\right){}^2 \left(-1+e^{2 i \pi 
      \left(a_1+a_4\right)} u'\right)}{\left(-1+e^{2 i \pi 
      \left(a_1-a_2\right)} u\right) \left(-1+e^{2 i \pi 
      \left(a_1+a_2\right)} u\right) \left(-1+\left(u'\right)^2\right)^2}\label{functions E}
\end{align}
\section{Simplifying the Painlev\'e connection coefficient }\label{app_simplifying}

Taking the logarithmic derivative of \eqref{core} and using

\begin{eqnarray}
\p_x\log{\hat{G}(x)}=\log{2\pi}-\pi x\cot{\pi x}
\end{eqnarray}
one obtains

\begin{eqnarray}
\p_x \log{\cm}=\sum_k\left(\log 2\pi -\pi(\omega_++\nu_k)\cot{\pi(\omega_++\nu_k)}\right)\p_x(\omega_++\nu_k)-\{\lambda\}\label{toapply}
\end{eqnarray}
Here and further $\{\lambda\}$ stays for the preceding terms with $\nu$ replaced by $\lambda$. Since $\sum_k\nu_k-\{\lambda\}=0$ we have
\begin{eqnarray}
\sum_k \log 2\pi\p_x(\omega_++\nu_k)-\{\lambda\}=0
\end{eqnarray}
Next, using the definition of $\omega_+$ \eqref{zpm} and the fact that $z_+$ satisfies \eqref{cubic} one can show that
\begin{eqnarray}
\sum_k -\pi \omega_+\cot\pi(\omega_++\nu_k)\p_x(\omega_++\nu_k)-\{\lambda\}=0
\end{eqnarray}
Therefore \eqref{toapply} is simplified down to
\begin{eqnarray}
\p_x\log{\cm}=\sum_k-\pi\nu_k\cot\pi(\omega_++\nu_k)\p_x(\omega_++\nu_k)-\{\lambda\}\label{toapplytrue}
\end{eqnarray}
Introduce the following difference operator 
\begin{eqnarray}
\delta_\al=\exp(\p/\p a^\al)-1
\end{eqnarray}
here and further $a^\al$ stays for any of the external/internal momentum. Note that
$\delta_\al$ annihilates periodic functions of $a^\al$ and also $\delta_\al a^\be=\delta^{\be}_\al$.

Apply $\delta_\al$ to \eqref{toapplytrue} and also set $x=a^\be$
\begin{eqnarray}
\delta_\al\p_\be\log{\cm}=\sum_k-\pi\delta_\al\nu_k\cot\pi(\omega_++\nu_k)\p_\be(\omega_++\nu_k)-\{\lambda\}
\end{eqnarray}
Here we used that $\delta_\al\nu_k$ is either 0 or 1. Therefore $\delta_\al\omega_+=0$ and $ \delta_\al\p_\be \nu_k=0$.

From the other hand, the logarithm of the function $\cm$ can be represented in the following form 
\begin{eqnarray}
\log{\cm}=\sum_{\al}a^\al P_\al+R
\end{eqnarray}
with all $P_\al$ and $R$ periodic. We therefore have
\begin{eqnarray}
\p_\be P_\al=\delta_\al\p_\be\log{\cm}=\sum_k-\pi\delta_\al\nu_k\cot\pi(\omega_++\nu_k)\p_\be(\omega_++\nu_k)-\{\lambda\}\label{toint}
\end{eqnarray}
Since $\delta_\al\nu_k$ and $\delta_\al\lambda_k$ are constants \eqref{toint} can be directly integrated to
\begin{eqnarray}
P_\al=\sum_k -\delta_\al\nu_k \log{\sin\pi(\omega_++\nu_k)}-\{\lambda\}
\end{eqnarray}
Note that we also have 
\begin{eqnarray}
\sum_{\al}a^\al\delta_\al\nu_k=\nu_k 
\end{eqnarray}
and the same for $\lambda_k$. Thus, we finally obtain
\begin{eqnarray}
\cm=\exp\left(\sum_{\al}a^\al P_\al+R\right)=\prod_k \frac{\sin\pi(\omega_++\nu_k)^{-\nu_k}}{\sin\pi(\omega_++\lambda_k)^{-\lambda_k}}e^{R}\label{simple}
\end{eqnarray}
This confirms that the non-periodic part of the formula \eqref{core} can be expressed in terms of the elementary functions. The remainder function $R$ probably can not be simplified in a similar fashion. 
\nocite{apsrev41Control}
\bibliographystyle{apsrev4-1}
\bibliography{bibfile,revtex-custom}
 
\end{document}